\documentclass{article}

\usepackage[preprint]{neurips_2024}

\usepackage[utf8]{inputenc} % allow utf-8 input
\usepackage[T1]{fontenc}    % use 8-bit T1 fonts
\usepackage{hyperref}       % hyperlinks
\usepackage{url}            % simple URL typesetting
\usepackage{booktabs}       % professional-quality tables
\usepackage{amsfonts}       % blackboard math symbols
\usepackage{nicefrac}       % compact symbols for 1/2, etc.
\usepackage{microtype}      % microtypography
\usepackage{xcolor}       
\usepackage{amsmath, amsthm}        %% for colors
\usepackage{extarrows}      %% for arrows
\usepackage{mathrsfs}       %% for \mathscr
\usepackage{tikz}           %% make graphics
\usetikzlibrary{shapes.geometric, arrows, calc, fadings, decorations.pathreplacing} %% more graphics

\tikzset{%
  >=latex, % option for nice arrows
  inner sep=0pt,%
  outer sep=2pt,%
  mark coordinate/.style={inner sep=0pt,outer sep=0pt,minimum size=3pt,
    fill=black,circle}%
}

\usepackage{textgreek} %% text greek
\usepackage{listings}  %% for code
\definecolor{dkgreen}{rgb}{0,0.6,0}
\definecolor{lviolet}{RGB}{231,154,255}
\definecolor{darkcyan}{RGB}{5, 55, 66}
\definecolor{dark_blue}{RGB}{3, 35, 42}
\title{Diffusion Map Autoencoder}
\author{%
  Julio Candanedo %\footnote{\texttt{jcandane@asu.edu}} %\thanks{Use footnote for providing further information about author (webpage, alternative address)---\emph{not} for acknowledging funding agencies.} 
  \\
  SparseTrace.ai , 2800 E. Enterprise Ave Ste 333, Appleton, WI 54913, USA \\
  \texttt{jcandane@asu.edu}, \texttt{julio@sparsetrace.ai}
  \\
}

\DeclareUnicodeCharacter{2208}{\ensuremath{\in}}
\usepackage{newunicodechar}
\newunicodechar{≫}{\ensuremath{\gg}}
\newunicodechar{≪}{\ensuremath{\ll}}
\newunicodechar{∑}{\ensuremath{\sum}}
\newunicodechar{ℓ}{\ensuremath{\ell}}
\newunicodechar{∀}{\ensuremath{\forall}}
\newunicodechar{∈}{\ensuremath{\in}}
\newunicodechar{ℝ}{\ensuremath{\mathbb{R}}}
\newunicodechar{≥}{\ensuremath{\geq}}
\newunicodechar{≤}{\ensuremath{\leq}}
\newunicodechar{−}{\ensuremath{-}}
\newunicodechar{†}{{\textdagger}}
\newunicodechar{→}{\ensuremath{\rightarrow}}
\newunicodechar{×}{\ensuremath{\times}}
\newunicodechar{ε}{\ensuremath{\varepsilon}}
\newunicodechar{μ}{\ensuremath{\mu}}
\newunicodechar{α}{\ensuremath{\alpha}}
\newunicodechar{β}{\ensuremath{\beta}}
\newunicodechar{∞}{\ensuremath{\infty}}

\usepackage{physics} % for \ket,\bra,\outerproduct etc. (optional)
\usepackage{microtype, bm}

\begin{document}
\maketitle

\begin{abstract}
Diffusion-Map-AutoEncoder (DMAE) pairs a diffusion-map encoder (using the Nyström method) with linear or RBF Gaussian-Process latent mean decoders, yielding closed-form inductive mappings and strong reconstructions.
\end{abstract}

\section{Introduction}

A central goal in manifold learning is to construct low–dimensional coordinates that preserve salient geometric or probabilistic structure of high–dimensional data, \cite{HintonSalakhutdinov2006}. Classic approaches differ mainly in (i) how they define inter-point relationships and (ii) how they turn those relationships into coordinates. The turn of the 21st century saw major and rapid development of nonlinear nonparametric methods: Kernel-Principal Component Analysis (K-PCA) by \cite{ScholkopfSmolaMuller1998_KPCA}, Locally Linear Embedding (LLE) by \cite{RoweisSaul2000_LLE}, Isomap by \cite{isomap_Tenenbaum2000, isomap_Balasubramanian2002}, Spectral-Eigenmaps by \cite{Belkin2003, Coifman2005}, GPLVM by \cite{Lawrence2005_GPLVM_JMLR}, t-SNE by \cite{MaatenHinton2008}, and UMAP by \cite{McInnes2018}.  
The above methods are fundamentally nonparametric: they compute coordinates for a \emph{fixed} dataset. Deploying them on novel points typically requires storing the training set (for OOS) and provides no native way to \emph{decode} (map latents back to ambients). This motivates \emph{parametric} variants that (a) keep the geometry of spectral/geodesic methods for the \emph{encoder} and (b) learn a \emph{decoder} (e.g., a kernel regressor or a linear map) for reconstruction and generation. Nyström OOS \cite{Nystrom2000} supplies a fast encoder for novel points; the representer theorem \cite{Scholkopf2001} yields kernel ridge (GP) decoders whose train–time and test–time formulas are closed form and consistent.
In this work we develop \emph{autoencoder} versions of diffusion maps in Diffusion-Map-AutoEncoder (DMAE), with principled OOS via Nyström, and the {decoder} is a kernel ridge (GP mean) map defined \emph{in latent space}.
This parametricization (i) enables accurate reconstruction and generative queries, (ii) decouples geometric choices in the encoder from approximation choices in the decoder, (iii) admits landmarks for scalability, and (iv) can be trained end–to–end (e.g., by refining decoder hyperparameters).

\subsubsection{Notation}

We use index, Einstein-summation, notation with all indices as \textbf{subscripts} of a symbol denoting a data-array.
Throughout, sample indices are $i,j,i',j'\in\mathbb{Z}_N$; ambient features $X,Y,X', Y'\in\mathbb{Z}_D$;
latent features $x,y,x',y'\in\mathbb{Z}_d$; novel/test samples $\alpha,\beta,\alpha',\beta'\in\mathbb{Z}_L$. Tensors are identified by index structure, e.g.,
the dataset (typically defined on ambient-features) is $R_{iX}$ with it's latent-features denoted as: $R_{ix}$. 
Furthermore, we name latent-feature pair-wise correlations by kernel-matrix $k_{ij} = \sum_x R_{ix}R^\top_{xj}$, and ambient-feature correlations by kernel-matrix $K_{ij} = R_{iX}R^\top_{Xj}$.
Some special-matrices are $\mathbf{1}_i$, which is a 1-index vector of entirely $1$s, with $\mathbf{1}_{ij}$ a matrix of entirely $1$s in the shape of the relevant indices, this is different from $I_{ij}$ which is the Kronecker-delta or identity-matrix (a diagonal 2-index matrix with $1$s on the diagonal). 
While \textbf{superscript} characters and numbers denote: exponents or if in text-mode they denote a label.

\section{Linear Autoencoder}\label{sec:lae}

\begin{figure}[!hbt]
    \centering
    \begin{tikzpicture}[node distance=2cm]
      \node (encode) [trapezium, trapezium angle=80, minimum width=20mm, minimum height=6mm, draw, thick, rotate=270] {\rotatebox{0}{LINEAR 1}};
      \node (decode) [trapezium, trapezium angle=80, minimum height=6mm, minimum width=20mm, draw, thick, rotate=90, yshift=-2.5cm ] {\rotatebox{180}{LINEAR 2}};
      \node (io) [ right of=decode] {$X$};
      \node (io2) [ left of=encode] {$X$};
      \draw (decode) -- node[above] {$x$} ++(encode);
      \draw (io2) -- (encode);
      \draw (io) -- (decode);
    \end{tikzpicture}
    \caption{Linear ``PCA'' autoencoder.}
    \label{fig:pca}
\end{figure}

Let's begin our discussion with PCA, suppose we have a dataset $R_{iX}$, assuming our dataset is centered at $\mu_X = \frac{1}{N}\sum_i R_{iX}$, our desire is dimensionality reduction, i.e. to determine $R_{ix}$. This map is provided by a \emph{loading-matrix} $V_{xX}$. 
Suppose we have novel-ambients $R_{\alpha X}$, and we wish to determine their embedding $R_{\alpha x}$, this can be achieved by projecting onto the PCA loadings (and using the same centering):
\begin{align}\label{pca_encoder}
    R_{\alpha x} &= \left(R_{\alpha X}-\mathbf{1}_\alpha\mu_X\right)\,V_{xX} = R_{\alpha X}V_{xX} + b_x \quad, \\
    R_{\alpha X} &= R_{\alpha x}V_{xX} + \mu_X \quad,
\end{align}
this equation is the Out-Of-Sample (OOS) extension to PCA. 
A notable feature is that all the columns of $V_{xX}$, over $x$ are orthogonal, $V_{x X}V^\top_{X x'} = I_{xx'}$ with $\left( \frac{1}{N}\sum_i \bar R_{iX}\bar R_{iY} \right)_{XY} V_{xY}=\lambda_x V_{xX}$.
An alternative and general view of this operation is a general matrix factorization $R_{iX} = R_{iy}M_{yX}$, with $M_{yX}$ not necessarily contained to be orthogonal. As made obvious in the bias ($b_x = - \mu_X V_{xX}$) in eq. \ref{pca_encoder}, this is a \emph{linear} layer, as is decoding.

\section{DMAE Theory}
\subsection{Standard Diffusion-map theory}

Let $R_{iX}\in\mathbb{R}^{N\times D}$ be the training dataset. We build a Gaussian-affinity in the ambient space
$K_{ij} = \exp\left(-\beta\,\|R_{iX}-R_{jX}\|^2\right)_{ij}$, with scale parameter: $\beta=\varepsilon^{-1}$.
To mitigate sampling-density bias we optionally use $\alpha$–normalization ($\alpha\in[0,1]$):
\begin{align}
K_i=\sum_{j}K_{ij},\qquad 
K^{(\alpha)}_{ij} \;=\; \frac{K_{ij}}{K_i^{\alpha}\,K_j^{\alpha}},\qquad
d_i=\sum_{j}K^{(\alpha)}_{ij}.
\end{align}
We consider two standard diffusion operators.

\emph{(i) Markov/random–walk (asymmetric)} operator is defined as:
$ P^{(\alpha)}_{ij} = d^{-1}_i K^{(\alpha)}_{ij}$. Let $(\lambda_x,\psi_x)$ be right–eigenpairs $P^{(\alpha)}_{ij}\psi_{jx} = \lambda_x\psi_{ix}$, ordered $1=\lambda_0\ge \lambda_1\ge\cdots$. The (time–$t\in\mathbb{R}$, an exponent) diffusion coordinates are:
\begin{align}
R_{ix} = \lambda_x^{t}\,\psi_{ix} \mathop{=}^{t\to 0} \psi_{ix}\qquad .
\end{align}

\emph{(ii) Symmetric} operator is defined by: $A^{(\alpha)}_{ij} = d^{-1/2}_i K^{(\alpha)}_{ij} d^{-1/2}_j$.
Let $(\lambda_x,u_x)$ satisfy $A^{(\alpha)}_{ij}u_{jx}=\lambda_x u_{ix}$ with $\lambda_0=1$ and $u_0\propto \sqrt d$. The symmetric diffusion coordinates are:
\begin{align}
R_{ix} = \lambda_x^{t}\,\frac{u_{ix}}{\sqrt{d_i}} \mathop{=}^{t\to 0} \frac{u_{ix}}{\sqrt{d_i}} \qquad .
\end{align}

In both cases the embeddings or latents, produce a linear-kernel (labeled a lower-case $k$, versus the uppercase ambient kernel $K_{ij}$), $k_{ij} = R_{ix}R^\top_{xj}$. In the limit $d\to D$, the latent linear-kernel is an exact factorization of the nonlinear-ambient kernel: $K_{ij} \propto k_{ij}$.

\subsection{Nyström Encoder}

Following work by \cite{Nystrom2000, Erichson2018}, given a novel ambient data-set batch $R_{\alpha X}$, we compute the test–train affinities and their $\alpha$–normalization:
\begin{align*}
K_{\alpha i}=\exp\!\big(-\beta\,\|R_{\alpha X}-R_{iX}\|^2\big),\qquad
K_\alpha=\sum_i K_{\alpha i},\qquad
K^{(\alpha)}_{\alpha i}=\frac{K_{\alpha i}}{K_\alpha^{\alpha} K_i^{\alpha}},\qquad
d_\alpha=\sum_i K^{(\alpha)}_{\alpha i}.
\end{align*}
We then may apply this to the two kinds of diffusion-maps:
\begin{align}
R_{\alpha x} &= \frac{K^{(\alpha)}_{\alpha i}}{d_\alpha}\,\frac{R_{ix}}{\lambda_x} & &\text{(Markov \& Symmetric).} 
\end{align}
These OOS formulas recover the training coordinates when $\alpha$ coincides with a training index (up to round-off), provided the same $(\beta,\alpha,t)$ are used.
In principle, a common neural-network (NN) can be student-teacher trained on diffusion map embeddings, \cite{Gal2019}. However, in this work we opt for a structurally similar sequential-NN to our Nyström encoder, this is shown in fig. \ref{fig:encoder}. Ultimately, the trainable hyper-parameters used are $\{\beta, R_{iX}, R_{ix}, \lambda_x\}$.
\begin{figure}[!hbt]
    \centering
    \begin{tikzpicture}[node distance=2cm]
      \node (encode2) [trapezium, trapezium angle=80, minimum width=30mm, minimum height=7mm, draw, thick, rotate=270] {\rotatebox{0}{LINEAR}};
      \node (encode) [trapezium, trapezium angle=40, minimum height=7mm, minimum width=30mm, draw, thick, rotate=270, yshift=-2.5cm ] {\rotatebox{0}{RBF}};
      \node (io) [ left of=encode] {$X$};
      \node (io2) [ right of=encode2] {$x$};
      \draw (encode) -- node[above] {$i$} ++(encode2);
      \draw (io2) -- (encode2);
      \draw (io) -- (encode);
    \end{tikzpicture}
    \caption{Diffusion map encoder architecture.}
    \label{fig:encoder}
\end{figure}

\subsection{Linear GP Decoder}

We may obtain the novel-ambients (from novel latents $R_{\alpha x}$) following the latent-mean Gaussian-Process-Interpolation (GPI) formula (also known as Kernel-Ridge, with uncertainty $\sigma^2$):
\begin{align*}
    R_{\alpha X} &= k_{\alpha i} \left( K_{ij} + \sigma^2 I_{ij} \right)^{-1}_{ij} R_{jX} \quad,
\end{align*}
or equivalently $\lim_{d\to D} k_{ij} = K_{ij}$ (ambient-kernel is equal to the latent kernel) in diffusion-maps. 
Crucially in the equation above with diffusion-map pieces, $K$ is the raw unnormalized Gaussian-affinity kernel, and the novel-training latent kernel $k_{\alpha i}(R_{\alpha x}, R_{ix})$ is also not normalized, such that conceptually if the novel-points were set to the training-points both kernels would cancel, mapping training ambients directly to the ``novel'' ambients.  
In diffusion maps the latent kernel is linear ($k_{ij} \propto R_{ix}R^\top_{xj}$), given this structure we may recast the formula: 
\begin{align}
    R_{\alpha X} &= R_{\alpha x} \underbrace{ R_{xi} \left( K_{ij} + \sigma^2 I_{ij} \right)^{-1}_{ij} R_{jX} }_{\text{weights}} \quad, \\
    R_{\alpha X} &= R_{\alpha x} \Theta_{xX} \quad.
\end{align}
Hence the matrix $\Theta_{xX}$ provides our Linear-decoding, it may be computed by the pieces listed above which are not needed beyond the training phase. In the LGP theory, this is our only trainable weight $\{\Theta_{xX}\}$.

% Requires: \usepackage{amsmath,amssymb}
\subsection{RBF (GP) decoder}

For a novel batch of latents \(R_{\alpha x}\in\mathbb{R}^{A\times d}\), define the \emph{latent} RBF kernel (unnormalized Gaussian affinity, mirroring the ambient kernel) with bandwidth $\varepsilon_{\text{lat}}>0$:
\begin{align*}
k_{\alpha i} =
\exp\!\left(-\tfrac{\lVert R_{\alpha x}-R_{ix}\rVert^2}{\varepsilon_{\text{lat}}}\right)
\in \mathbb{R}^{A\times N},
\qquad k_{ij} = \exp\left(-\tfrac{\lVert R_{ix}-R_{jx}\rVert^2}{\varepsilon_{\text{lat}}}\right) \quad.
\end{align*}
Treating each ambient coordinate (each column of ($R_{iX}$) as an independent Gaussian Process with the same kernel and i.i.d. Gaussian observation noise $\sigma^2$, the posterior mean (equivalently, Kernel Ridge Regression) yields the decoder:
\begin{align}
R_{\alpha X} = k_{\alpha i}\underbrace{\left(k_{ij}+\sigma^2 I_{ij}\right)^{-1}\,R_{jX} }
\quad.
\label{eq:gp-decoder}
\end{align}
It is convenient to precompute the \emph{decoder weight matrix}:
\begin{align}
S_{iX} := \left(k_{ij}+\sigma^2 I_{ij}\right)^{-1} R_{jX}, \qquad .
\label{eq:weights-S}
\end{align}
If the ``novel'' latents coincide with the training latents ($\alpha=i$), then:
\begin{align*}
R_{iX} = k_{ij}\,\left(k_{jk}+\sigma^2 I_{jk}\right)^{-1} R_{kX},
\end{align*}
which equals $R_{iX}$ in the noise–free limit $\sigma^2\to 0$ (and is a slight shrinkage otherwise).
Choosing the \emph{linear} latent kernel $k(z,z')=z^\top z'$ gives $k_{\alpha i}=R_{\alpha x}R_{xi}$ and $k_{ij}=R_{ix}R_{xj}$. Plugging into \eqref{eq:gp-decoder} yields
\begin{align}
R_{\alpha X} =
R_{\alpha x}\underbrace{\,\left[\,R_{xi}\left(R_{ix}R_{xj}+\sigma^2 I_T\right)^{-1}R_{jX}\,\right]}_{\Theta_{xX}}
= R_{\alpha x}\Theta_{xX},
\label{eq:linear-kernel-form}
\end{align}
which is the kernel-form of the linear (ridge) decoder. Equivalently, via the Woodbury-identity, the \emph{primal} ridge form is:
\begin{align}
\Theta_{xX} =
\left(R_{xi}R_{ix}+\sigma^2 I_{xx} \right)^{-1} R_{xi} R_{iX} \quad.
\label{eq:linear-primal}
\end{align}
We keep the latent kernel \emph{unnormalized} (pure Gaussian affinity), consistent with the diffusion-map construction in ambient space. The only new hyperparameter is $\varepsilon_{\text{lat}}$, for which we use a median-distance heuristic on $\{R_{ix}\}$. Training cost is dominated by a single Cholesky of $k+\sigma^2 I$ (i.e., $O(N^3)$); test-time decoding is the matrix product $k_{\alpha i}S_{iX}$ (i.e., $O(AND)$).
The hyperparameter weights for our RBF-GP decoder are: $\{\varepsilon^\text{lat}, R_{ix}, S_{iX} \}$

\subsection{Autoencoder}

\begin{figure}[!hbt]
    \centering
    \resizebox{0.42\linewidth}{!}{%
      \begin{tikzpicture}[node distance=2cm, scale=0.50]
      \node (encode) [trapezium, trapezium angle=40, minimum height=7mm, minimum width=30mm, draw, thick, rotate=270, yshift=-2.5cm ] {\rotatebox{0}{RBF}};
      \node (encode2) [trapezium, trapezium angle=80, minimum width=30mm, minimum height=7mm, draw, thick, rotate=270] {\rotatebox{0}{LINEAR}};
      \node (decode) [trapezium, trapezium angle=120, minimum width=30mm, minimum height=7mm, draw, thick, rotate=270, yshift=3.0cm] {\rotatebox{0}{LINEAR}};
      \node (io11)  [ left of=encode ] {$X$};
      \node (io22)  [ right of=decode ] {$X$};
      %%%%%%%%%%%
      \draw (encode)  -- node[above] {$i$} ++(encode2);
      \draw (encode2) -- node[above] {$x$} ++(decode);
      \draw (io11) -- (encode);
      \draw (decode) -- (io22);
    \end{tikzpicture}
    }\quad\qquad
    \resizebox{0.50\linewidth}{!}{%
    \begin{tikzpicture}[node distance=2cm, scale=0.45]
        \node (encode) [trapezium, trapezium angle=40, minimum height=7mm, minimum width=30mm, draw, thick, rotate=270, yshift=0.0cm ] {\rotatebox{0}{RBF}};
        \node (io) [left of=encode] {$X$};
        \node (encode2) [trapezium, trapezium angle=80, minimum width=30mm, minimum height=6.5mm, draw, thick, rotate=270, above of=encode ] {\rotatebox{0}{LINEAR}};
        \node (decode2) [trapezium, trapezium angle=40, minimum width=30mm, minimum height=6.5mm, draw, thick, rotate=90, below of=encode2 ] {\rotatebox{180}{RBF}};
        \node (decode3) [trapezium, trapezium angle=80, minimum width=30mm, minimum height=6.5mm,  draw, thick, rotate=90, below of=decode2 ] {\rotatebox{180}{LINEAR}};
        \node (io2) [ right of=decode3] {$X$};
      \draw (io) --node[above] {} ++(encode);
      \draw (encode)  -- node[above] {$i$} ++(encode2);
      \draw (encode2) -- node[above] {$x$} ++(decode2);
      %\draw (decode)  -- node[above] {$\alpha$} ++(decode2);
      \draw (decode2) -- node[above] {$j$} ++(decode3);
      \draw (decode3)  --node[above] {} ++(io2);
      %%%
    \end{tikzpicture}}
    \caption{DMAE Architectures: Linear Decoder (left) and DMAP (``RBF'')-layer decoder (right) .}
    \label{fig:autocoder}
\end{figure}

Finally, combining all our pieces we have $\{ \beta, R_{ix}, R_{iX}, \lambda_x \}$ parameters from our encoder, and $\{\Theta_{xX}\}$ (linear) or $\{\varepsilon^\text{lat}, R_{ix}, S_{iX}\}$ (RBF) parameters for our decoder, these are the \emph{weights} of our autoencoder. Taking their union we can the trainable weights to two kinds of autoencoders (linear and RBF version, with their hyperparameter count):
\begin{align*}
    \text{RBF } &\{ \beta, \varepsilon^\text{lat}, \lambda_x, R_{ix}, R_{iX}, S_{iX}  \} \sim \mathcal{O}(1 + 1 + d + Nd + ND + ND) = \mathcal{O}(ND), \\
    \text{Linear } &\{ \beta, \lambda_x, \Theta_{xX}, R_{ix}, R_{iX} \}\sim \mathcal{O}(1 + d + Dd + Nd + ND) = \mathcal{O}(ND).
\end{align*}
However, if we use our decoder to decode our latents back to our dataset $R_{ix}\to R_{iX}$, these parameters can be computed as needed, leading to the following counts:
\begin{align*}
    \text{RBF } &\{ \beta, \varepsilon^\text{lat}, \lambda_x, R_{ix}, S_{iX}  \} \sim \mathcal{O}(1 + 1 + d + Nd + ND) = \mathcal{O}(ND), \\
    \text{Linear } &\{ \beta, \lambda_x, \Theta_{xX}, R_{ix} \}\sim \mathcal{O}(1 + d + Dd + Nd) = \mathcal{O}(Nd).
\end{align*}
Recall our original dataset $R_{iX}$, has $\mathcal{O}(ND)$ parameters, however this DMAE can also be applied to novel samples $R_{\alpha X}$, and when cast into Sequential-NN form can be finetuned or optimized with novel data (for fixed parameter count).

\section{Experiments}

\subsection{Swiss-Roll}

\begin{figure}[hbt!]
    \centering
    \includegraphics[width=0.95\linewidth]{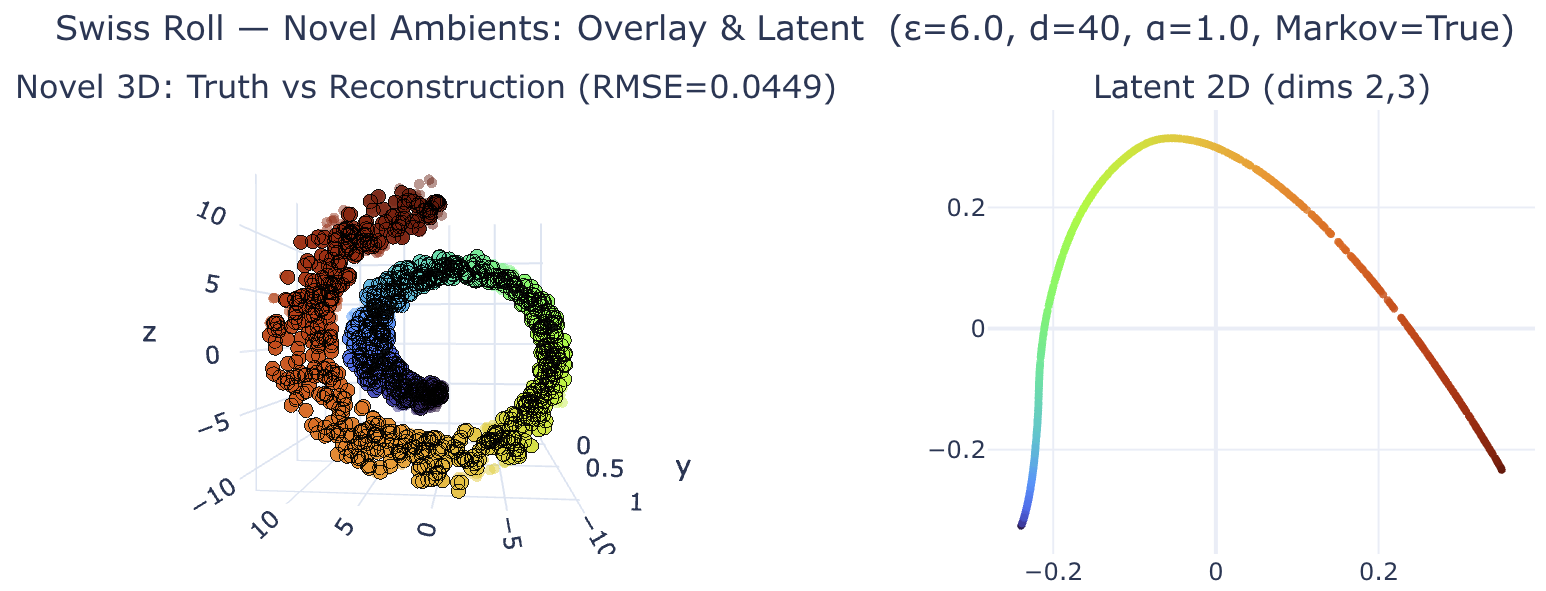}
    \caption{Above is an example of the Swiss-roll data set in the ambient (left, with both novel original and reconstructions) and latent coordinates (right).}
    \label{fig:basic_swiss_image}
\end{figure}

We evaluate our proposed Diffusion-Map-Autoencoder (DMAE) on the canonical Swiss-roll manifold (popularized by Isomap, \cite{isomap_Tenenbaum2000}) to illustrate: (i) qualitative behavior of the encoder/decoder, (ii) the effect of latent dimension $d$, (iii) sensitivity to the ambient bandwidth $\varepsilon$, and (iv) robustness to observation noise. Unless stated otherwise we use $\alpha=1.0$ (co-normalization), a Markov (random-walk) diffusion operator, diffusion time $t=0.5$, and ridge/GP regularization $\sigma^2=10^{-4}$; all experiments are in double-precision (\texttt{float64}). 
For each plot we fix a single train/test split (2,000 train, 1,200 test) and keep it constant across the corresponding sweep.
Figure~\ref{fig:basic_swiss_image} gives a representative example of our Swiss-roll: the ambient 3D point cloud (left) overlays novel test inputs and their reconstructions, while the latent 2D (first two nontrivial embedding vectors) embedding (right) exposes the learned intrinsic geometry. The DMAE encoder yields a coherent 2D parameterization of the roll, and both decoders (linear and RBF/GP) produce visually faithful reconstructions.

\begin{figure}[hbt!]
    \centering
    \includegraphics[width=0.98\linewidth]{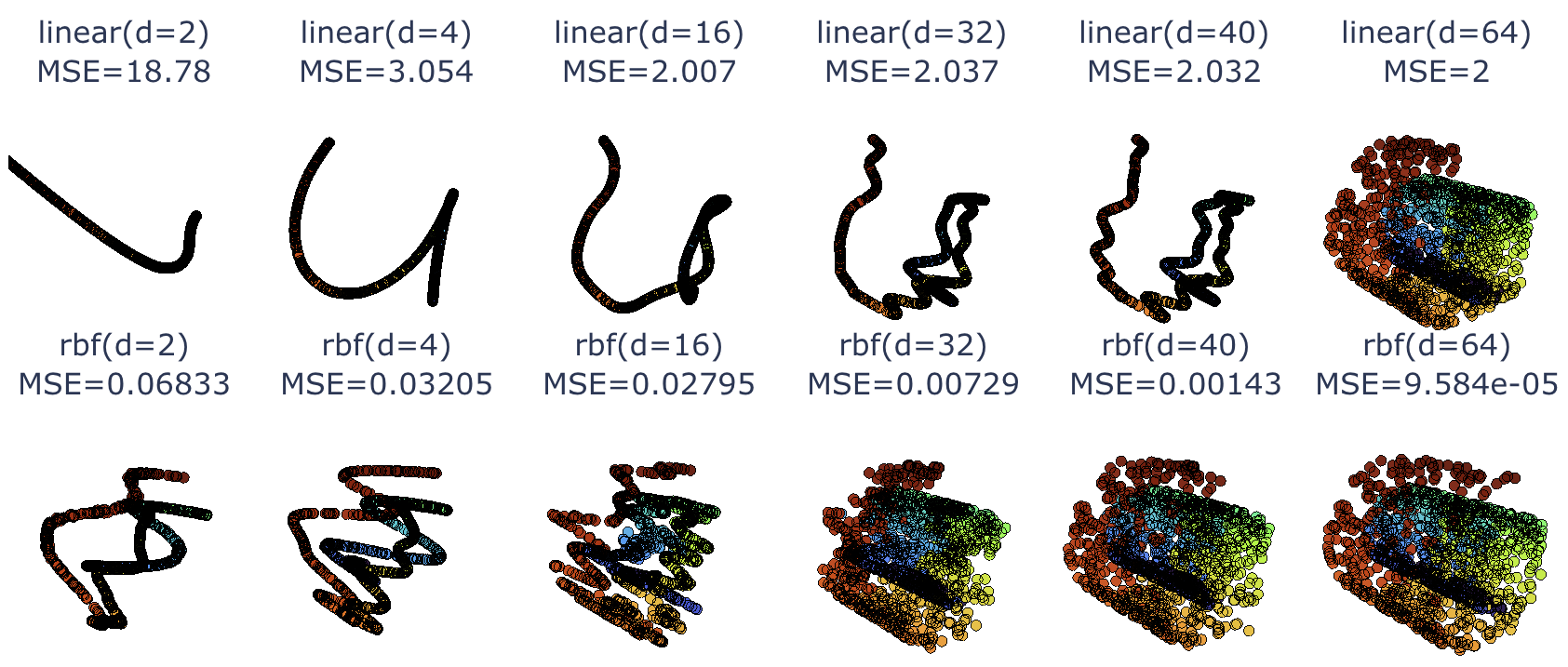}
    \caption{We have two series of 3D plots showing the reconstruction of the Swiss-roll. (top) is done with a linear-GP decoder for a series of 6 latent dimensions. (bottom) is done with an RBF-layer.}
    \label{fig:deps}
\end{figure}

\begin{figure}[hbt!]
    \centering
    \includegraphics[width=0.98\linewidth]{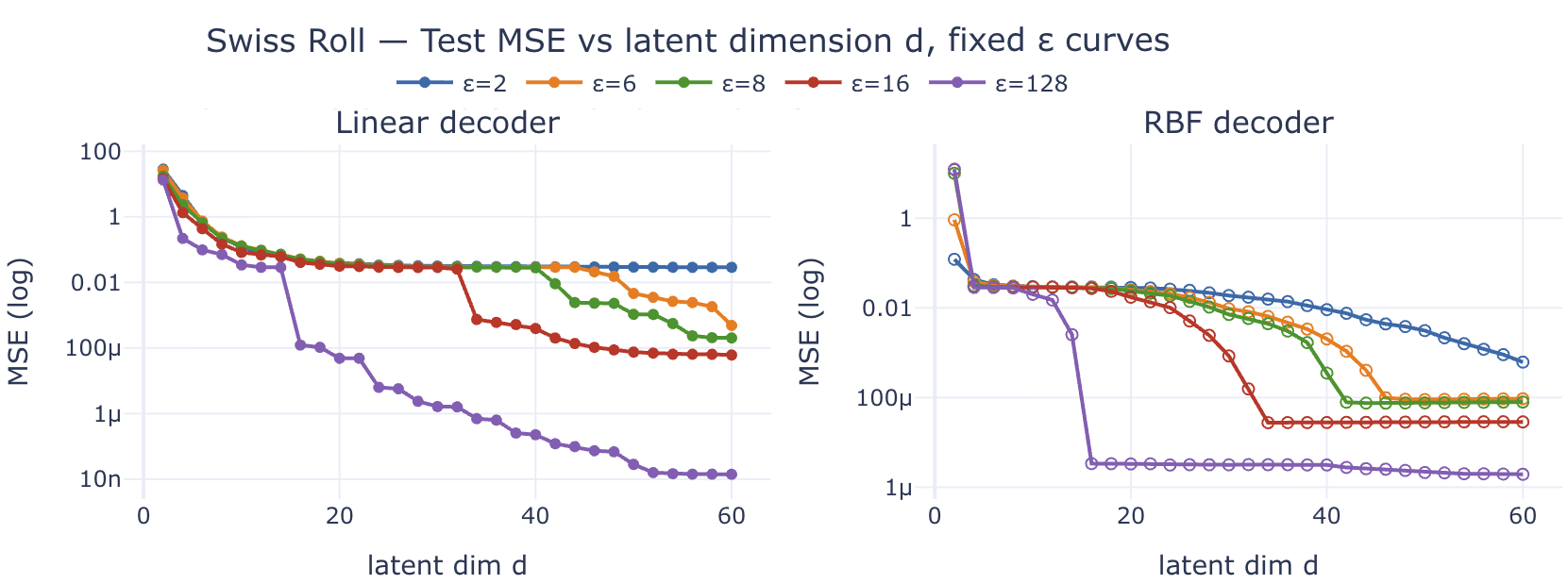}
    \caption{We compare the linear and RBF decoder layers, by the MSE as a function of latent dimension and scale parameter $\varepsilon$.}
    \label{fig:mse_swiss}
\end{figure}

Figure~\ref{fig:deps} stacks 2x6 3D panels of reconstructions only: the top row uses the \emph{linear} decoder and the bottom row the \emph{RBF/GP} decoder, each across $d\in\{2,4,16,32,40,64\}$ with a fixed ambient bandwidth $\varepsilon=6$ (a known length scale to capture intrinstic structure). As $d$ increases, both decoders improve rapidly and then saturate once the intrinsic degrees of freedom are captured (dependent on our scale $\beta,\varepsilon$). The RBF/GP decoder is visibly stronger at small $d$ (less underfitting), while the linear decoder can exhibit slight geometric distortions at low $d$ that diminish as capacity grows.
To quantify these trends, Fig.~\ref{fig:mse_swiss} plots test MSE as a function of $d$ for a grid of ambient bandwidths $\varepsilon\in\{2,6,8,16,128\}$ (for left: linear, right: RBF/GP). Two consistent patterns emerge. 
(i) For fixed $\varepsilon$, error decreases steeply with $d$ and then plateaus once the model is sufficiently expressive. 
(ii) For fixed $d$, the choice of $\varepsilon$ is crucial: very small $\varepsilon$ under-connects the graph (a highly nonlinear solution, requiring a high rank linear decoder), whereas very large $\varepsilon$ yields a linear like (low-rank) model. The RBF/GP decoder is more robust across a wider range of $\varepsilon$ values, reflecting its nonlinear interpolation in latent space.

\begin{figure}[hbt!]
    \centering
    \includegraphics[width=0.85\linewidth]{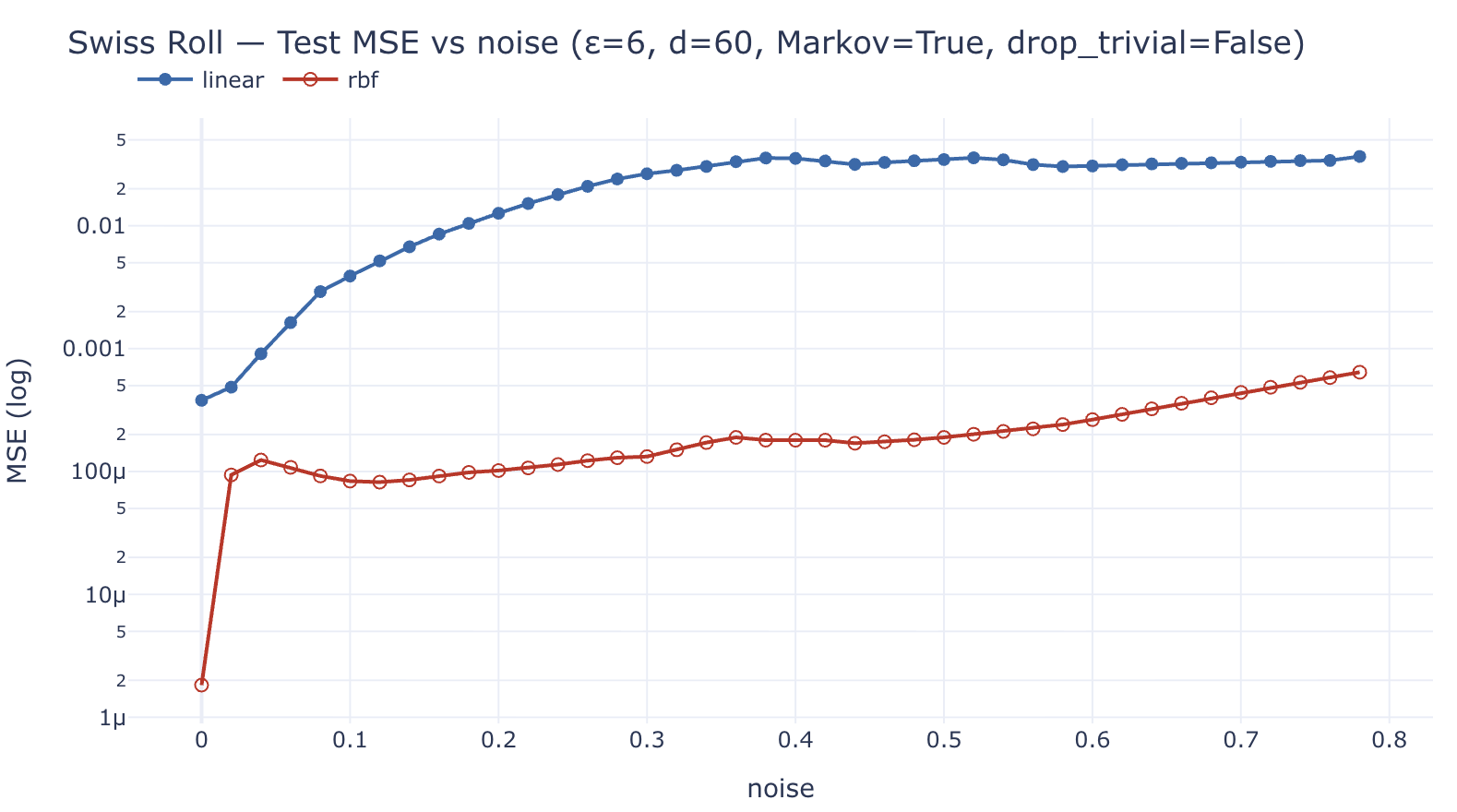}
    \caption{We determine the MSE as a function of noise of the original Swiss-roll: linear and RBF decoder layers. }
    \label{fig:swiss_noise}
\end{figure}

Our final test is presented in Figure~\ref{fig:swiss_noise} reports test MSE as a function of additive Gaussian noise level on the inputs (both train and test are generated at the same $\sigma$), with $d$ and $\varepsilon$ fixed. As expected, error increases with $\sigma$. Across all noise levels, the RBF/GP decoder maintains a consistent gap over the linear decoder—its kernel smoothing in latent space damps high-frequency perturbations, yielding lower MSE. The linear decoder remains competitive at low noise when $d$ is large, but degrades faster as noise increases.

On Swiss-roll, (1) increasing $d$ helps until the intrinsic structure is captured, after which returns diminish; (2) there is a clear sweet spot in $\varepsilon$ balancing bias and variance; and (3) the RBF/GP decoder consistently improves inductive generalization, particularly in low-$d$ or noisy regimes. These observations are consistent with the theory: the linear decoder solves a ridge regression in diffusion coordinates, while the RBF/GP decoder implements the posterior mean of a Gaussian Process in latent space, providing a smoother, more flexible inductive map.

\subsection{MNIST}

\begin{figure}[hbt!]
    \centering
    \includegraphics[width=0.85\linewidth]{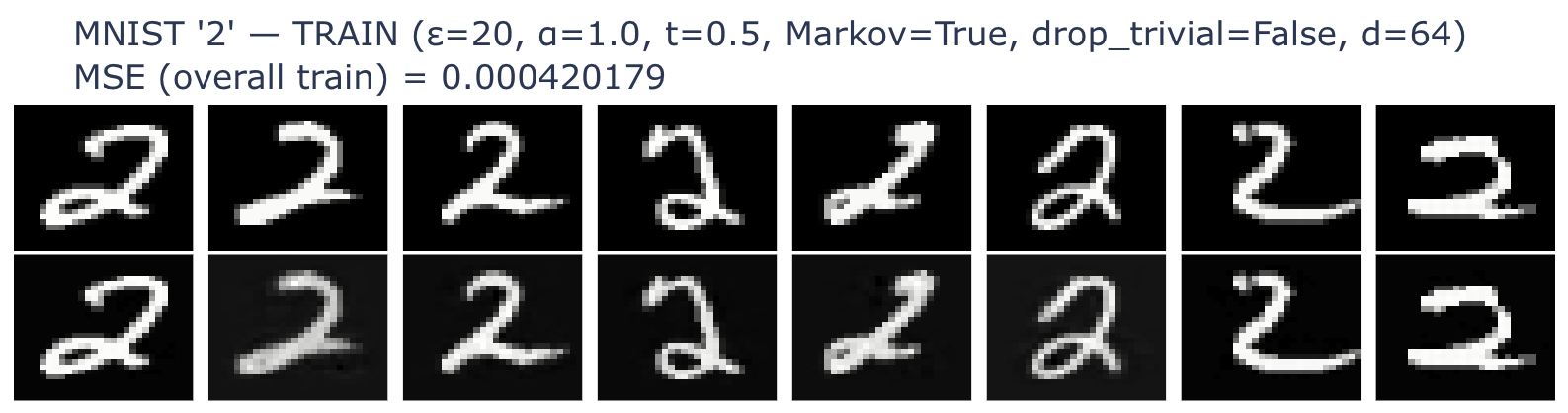}\\[4pt]
    \includegraphics[width=0.85\linewidth]{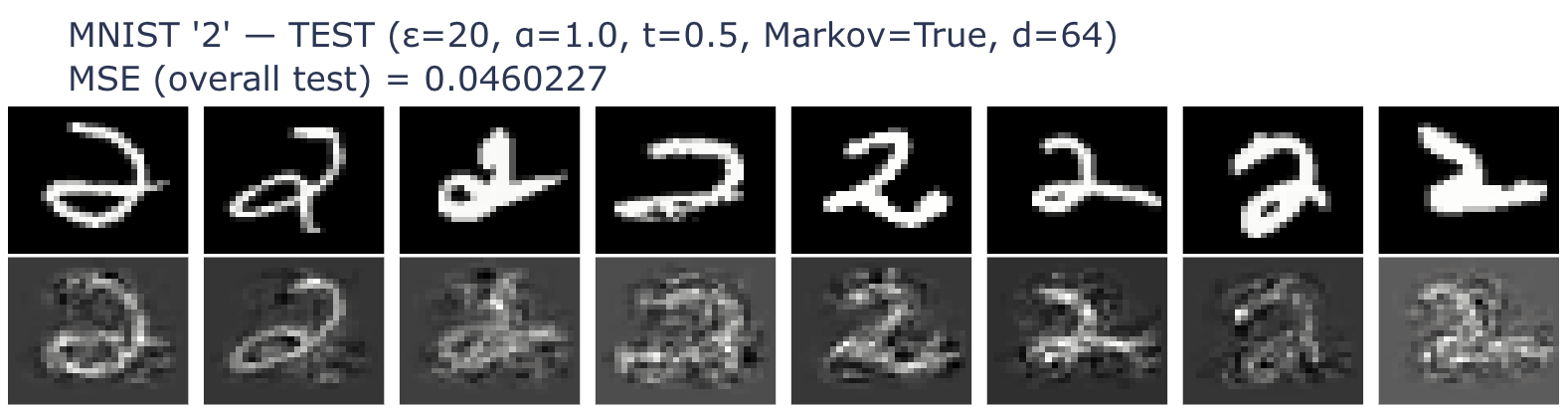}
    \caption{Grids of MNIST-2 images (top rows: truth; bottom rows: reconstructions) using the DMAE with RBF/GP decoder at $d=64$. Top: training samples. Bottom: novel test samples.}
    \label{fig:mnist_recon}
\end{figure}

\begin{figure}[hbt!]
    \centering
    \includegraphics[width=0.75\linewidth]{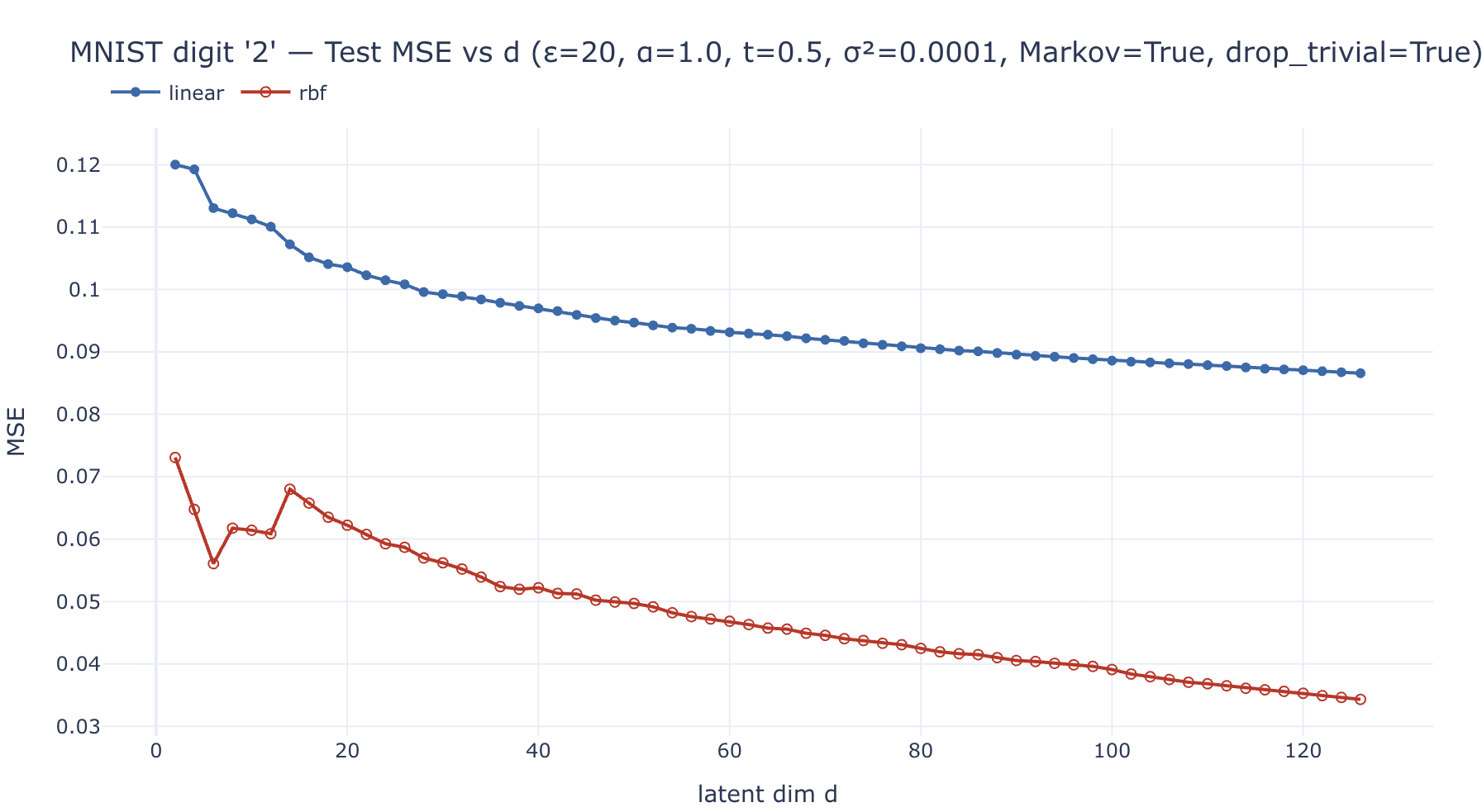}
    \caption{Test MSE vs.\ latent dimension $d$ on MNIST-2 with $\varepsilon=20$. We compare the linear decoder (ridge regression in diffusion coordinates) and the RBF/GP decoder (posterior mean with an RBF kernel in latent space).}
    \label{fig:mnist_mse_curve}
\end{figure}

We next study a real image manifold using MNIST (\cite{LeCunCortesMNIST}). To control intra-class variation we restrict to the digit ``2'' (MNIST-2), normalize pixels to $[0,1]$, and use a fixed split of 2,000 training and 1,200 test images. Unless noted otherwise we set the ambient kernel bandwidth to $\varepsilon=20$, use $\alpha=1.0$ (co-normalization) with a Markov (random-walk) operator, diffusion time $t=0.5$, and ridge/GP regularization $\sigma^2=10^{-4}$; computations are in double-precision, \texttt{float64}.
Figure~\ref{fig:mnist_recon} shows $2\times 8$ grids illustrating input images (top rows) and DMAE reconstructions (bottom rows). The top panel contains samples drawn from the training set; the bottom panel uses novel test samples. We use the RBF/GP decoder with $d=64$ latent dimensions (to match the Swiss-roll setting in Fig.~\ref{fig:deps}). Training reconstructions are near-perfect, as expected with a flexible decoder, while test reconstructions maintain high fidelity—capturing stroke thickness and curvature with only mild smoothing of fine details.
To quantify capacity trends, Fig.~\ref{fig:mnist_mse_curve} plots the test MSE as a function of the latent dimension $d$ for both the linear and RBF/GP decoders, with $\varepsilon$ fixed at $20$ (one train/test split held constant across the sweep). Error decreases with $d$ and then plateaus once the dominant modes of variation are captured (with a plateau likely directly related to the scale-factor $\varepsilon$). The RBF/GP decoder dominates at small to moderate $d$ (lower bias via nonlinear interpolation in latent space), while the linear decoder narrows the gap as $d$ grows. At very large $d$, slight flattening or even a gentle uptick can occur for the linear curve if regularization is too small; increasing $\sigma^2$ mitigates this.
On MNIST-2, the DMAE provides competitive reconstruction accuracy with a simple, interpretable pipeline. The RBF/GP decoder offers stronger inductive generalization at modest latent dimension. In all cases we reconstruct via encode$\rightarrow$decode in latent space (not a direct ambient mapping), ensuring a fair comparison between decoders.

% --- Figures (renamed labels to avoid duplication) ---

\section{Conclusion}

We introduced a Diffusion–Map Autoencoder (DMAE) that pairs a nonparametric diffusion–map encoder with two lightweight decoders: (i) a linear ridge map in diffusion coordinates, and (ii) an RBF/GP decoder that is exactly the posterior mean of a Gaussian Process in latent space. The result is an inductive, out-of-sample autoencoder with no learned convolutions and with clear knobs—ambient bandwidth $\varepsilon$, normalization $\alpha$, diffusion time $t$, and latent dimension $d$—that directly control the bias–variance trade-off.
On Swiss–roll, qualitative overlays and quantitative curves show that reconstruction error decreases sharply with $d$ and then saturates, and that performance is highly sensitive to $\varepsilon$: too small under–connects the graph and leads to a highly nonlinear and high-rank linear model. Hence paradoxically, the parameter count increases with increased nonlinearly ($\varepsilon\to 0$). Across $\varepsilon$ and under input noise, the RBF/GP decoder consistently dominates at small–to–moderate $d$, while the linear decoder becomes competitive once $d$ is sufficiently large. On MNIST (digit ``2''), DMAE attains low test MSE with simple settings: training reconstructions are near–perfect and test reconstructions preserve stroke geometry with only mild smoothing. The RBF/GP decoder again delivers better inductive generalization for a fixed $d$.

Compared to standard AEs/VAEs, DMAE offers: (i) an explicit geometric encoder grounded in spectral theory; (ii) a principled GP interpretation of the nonlinear decoder (yielding a natural uncertainty proxy via predictive variance); and (iii) strong sample efficiency on small training sets through kernel interpolation in latent space. Practically, the GP decoder achieves higher compression than the linear map—reaching a target MSE with substantially smaller $d$—which is attractive when storage or downstream compute is constrained.
The RBF/GP decoder requires a Cholesky of a $N\times N$ kernel (time $\mathcal{O}(N^3)$, memory $\mathcal{O}(N^2)$), limiting very large $N$ unless approximations are used. Performance hinges on sensible choices of $\varepsilon$, $\alpha$, and $t$; poorly tuned bandwidths can degrade both decoders, especially the linear map at low $d$. Our MNIST study considers a single class; multi-class and higher-resolution datasets are left for future work.

Beyond stationary RBFs, deep-kernel learning could improve low–$d$ accuracy while retaining an inductive, closed–form decoder.  
Additionally VAE-style DMAE, a probabilistic variant that treats a small-variance Gaussian around the diffusion coordinates and and optimizes an ELBO with a Gaussian or Bernoulli likelihood would provide calibrated generative modeling while preserving the geometric encoder; the GP predictive variance can regularize the recon term or guide $\beta$–annealing.  
Replacing diffusion maps with Isomap (geodesic distances) or other graph–geometric embeddings offers an interesting encoder family; the same linear and GP decoders apply, yielding ``Isomap–AE'' variants. 
Inducing-point GPs, kernel interpolation, conjugate-gradient solvers with fast MVMs, or multi-scale $\varepsilon$ schedules can extend DMAE to much larger $N$.
A diffusion (or geodesic) encoder coupled with a GP mean decoder is a simple, interpretable recipe that delivers competitive reconstructions on nonlinear manifolds, with clear, tunable controls for geometry and smoothness and a straightforward path to probabilistic, VAE-style extensions.

%\newpage
\bibliographystyle{fancybib}
\bibliography{refs}

\end{document}